\documentclass[12pt]{iopart}

\usepackage{iopams}
\usepackage{amssymb}
\eqnobysec

\begin{document}

\title[]{The hypergeneralized Heun equation in QFT in curved space-times}

\author{D. Batic}
\address{Department of Mathematics, University of Los Andes, AA 4976 Bogota, Colombia}
\ead{dbatic@uniandes.edu.co}
\author{M. Sandoval}
\address{Department of Mathematics, University of Los Andes, AA 4976 Bogota, Colombia}
\ead{ma-sando@uniandes.edu.co}

\begin{abstract}
In this article we show for the first time the role played by the
hypergeneralized Heun equation (HHE) in the context of Quantum Field
Theory in curved space-times. More precisely, we find suitable
transformations relating the separated radial and angular parts of a
massive Dirac equation in the Kerr-Newman-deSitter metric to a HHE.

\end{abstract}

\pacs{04.62.+v, 04.70.-s}
\submitto{\JPA}
\maketitle

\section{Introduction}\label{sec:1}
The hypergeneralized Heun equation (HHE) is a second order
differential equation of the form
\begin{equation}\label{HHE}
\fl y^{''}(z) +\sum_{i=0}^{4}\frac{1-\mu_{i}}{z-z_i}~
y^{'}(z)+\frac{\beta_{0}+\beta_{1}z+\beta_{2}z^{2}+\beta_{3}z^{3}}{\prod\limits_{i=0}^{4}(z-z_i)}~y(z)=0.
\end{equation}
where $z_0=0$, $z_1=1$, $z_2$, $z_3$,
$z_4\in\mathbb{C}\backslash\{0,1\}$ and $\mu_{0},\cdots\mu_4$,
$\beta_{0},\cdots,\beta_{3}$ are arbitrary complex numbers.
Moreover, $z_i$ is a simple singularity with exponents
$\{0,\mu_{i}\}$ for each $i=0,\cdots,4$ and $\infty$ is also a
simple singularity. To underline the importance of equation
(\ref{HHE}) we recall that it represents a generalization of the
equation studied by Schmidt in \cite{Schmidt} and it contains the
generalized Heun equation \cite{sch,batic} as a confluent special
case. The aim of our work is to provide the first example where the
HHE arises in physics. In particular, we are interested in the role
played by (\ref{HHE}) in QFT theory in curved space-times.\\

\section{The Dirac equation in the Kerr-Newman-deSitter metric}\label{sec:2}
In Boyer-Lindquist coordinates $(t,r,\vartheta,\varphi)$ with $r>0$,
$0\leq\vartheta\leq\pi$, $0\leq\varphi<2\pi$ the
Kerr-Newman-deSitter metric is \cite{kan}
\[
\fl g^{\mu\nu}=\left(
\begin{array}{cccc}
\displaystyle\frac{\Xi^{2}}{\Sigma}\left[\frac{(r^{2}+a^{2})^{2}}{\Delta_{r}}-\frac{a^{2}\sin^{2}\vartheta}{\Delta_{\vartheta}}\right]&0&0&\displaystyle\frac{a\Xi^{2}}{\Sigma}\left[\frac{r^{2}+a^{2}}{\Delta_{r}}-\frac{1}{\Delta_{\vartheta}}\right]\\
0&-\displaystyle\frac{\Delta_{r}}{\Sigma}&0&0\\
0&0& -\displaystyle\frac{\Delta_{\vartheta}}{\Sigma} &0\\
\displaystyle\frac{a\Xi^{2}}{\Sigma}\left[\frac{r^{2}+a^{2}}{\Delta_{r}}-\frac{1}{\Delta_{\vartheta}}\right]&0&0&-\displaystyle\frac{\Xi^{2}}{\Sigma\sin^{2}\vartheta}\left[\frac{1}{\Delta_{\vartheta}}-\frac{a^{2}\sin^{2}\vartheta}{\Delta_{r}}\right]\\
\end{array}
\right)
\]
with
\[
\fl \Sigma:=\Sigma(r,\theta)=r^2+a^2\cos^{2}\theta, \qquad
\Delta_r:=\Delta_r(r)=(r^{2}+a^{2})\left(1-\frac{\Lambda}{3}
r^{2}\right)-2Mr+Q^{2},
\]
\[
\fl
\Delta_{\vartheta}:=\Delta_{\vartheta}(\vartheta)=1+\frac{\Lambda}{3}a^{2}\cos^{2}\vartheta,\qquad
\Xi=1+\frac{\Lambda}{3}a^{2}
\]
where $\Lambda$ is the cosmological constant and $M$, $a$ and $Q$
are the mass, the angular momentum per unit mass and the charge of
the black hole, respectively. In what follows we assume that the
function $\Delta_r$ has four distinct zeros. According to Penrose
and Rindler \cite{pen} the Dirac equation for two spinors $P^{A}$
and $Q^{A}$ is given by
\begin{equation}\label{Dirac_01}
\fl \nabla_{AA^{'}}P^{A}+\frac{\rmi
m_{e}}{\sqrt{2}}~\overline{Q}_{A^{'}}=0,\qquad
\nabla_{AA^{'}}Q^{A}+\frac{\rmi
m_{e}}{\sqrt{2}}~\overline{P}_{A^{'}}=0
\end{equation}
where we used Planck units $\hbar=c=G=1$. Furthermore,
$\nabla_{AA^{'}}$ is the symbol for covariant differentiation and
$m_{e}$ is the particle mass. The Dirac equation in the
Kerr-Newman-deSitter geometry was computed and separated by Khanal
\cite{kan}. If we make the following ansatz for the spinor $\Psi$
\[
\fl \Psi(t,r,\vartheta,\varphi)=\left( \begin{array}{c}
               +P^{0}(t,r,\vartheta,\varphi)\\
               +P^{1}(t,r,\vartheta,\varphi)\\
     +\overline{Q}^{1^{'}}(t,r,\vartheta,\varphi)\\
     -\overline{Q}^{0^{'}}(t,r,\vartheta,\varphi)
\end{array}\right)=\rme^{-\rmi\omega t}\rme^{\rmi\widehat{k}\varphi}\left( \begin{array}{c}
               F_{1}(r,\vartheta)\\
               F_{2}(r,\vartheta)\\
               G_{1}(r,\vartheta)\\
               G_{2}(r,\vartheta)
\end{array} \right),\quad\widehat{k}=k+\frac{1}{2},
\]
with $\omega\in\mathbb{R}$ and $k\in\mathbb{Z}$ the energy and the
azimuthal quantum number of the particle,respectively the equations
in (\ref{Dirac_01}) lead to the following coupled linear system of
first order PDEs
\begin{eqnarray}
\fl
&&\mathcal{D}_{0}F_{1}+\sqrt{\Delta_{\vartheta}/2}~\mathcal{L}_{1/2}F_{2}=
\frac{m_{e}}{\sqrt{2}}(\rmi r+a\cos\vartheta)G_{1},\label{E1}\\
\fl
&&\Delta_{r}\mathcal{D}^{\dag}_{1/2}F_{2}-\sqrt{2\Delta_{\vartheta}}~\mathcal{L}^{\dag}_{1/2}F_{1}=
-\sqrt{2}m_{e}(\rmi r+a\cos\vartheta)G_{2},\label{E2}\\
\fl
&&\mathcal{D}_{0}G_{2}-\sqrt{\Delta_{\vartheta}/2}~\mathcal{L}^{\dag}_{1/2}G_{1}=
\frac{m_{e}}{\sqrt{2}}(\rmi r-a\cos\vartheta)F_{2},\label{E3}\\
\fl
&&\Delta_{r}\mathcal{D}^{\dag}_{1/2}G_{1}+\sqrt{2\Delta_{\vartheta}}~\mathcal{L}_{1/2}G_{2}=-\sqrt{2}m_{e}(\rmi
r-a\cos\vartheta)F_{1}\label{E4}
\end{eqnarray}
where
\begin{eqnarray}
\fl
&&\mathcal{D}_{0}=\displaystyle\frac{d}{dr}+\rmi\frac{\Xi K(r)}{\Delta_{r}},
\qquad K(r)=a\widehat{k}-\omega(r^{2}+a^{2}),\label{1}\\
\fl
&&\mathcal{D}_{1/2}=\displaystyle\frac{d}{dr}+\rmi\frac{\Xi
K(r)}{\Delta_{r}}+\frac{1}{2\Delta_{r}}\frac{d\Delta_{r}}{dr},\qquad
\mathcal{D}^{\dag}_{1/2}=\displaystyle\frac{d}{dr}-\rmi\frac{\Xi K(r)}{\Delta_{r}}+
\frac{1}{2\Delta_{r}}\frac{d\Delta_{r}}{dr}, \label{3}\\
\fl
&&\mathcal{L}_{1/2}=\displaystyle\frac{d}{d\vartheta}+\frac{\Xi H(\vartheta)}{\Delta_{\vartheta}}+
\frac{1}{2\sqrt{\Delta_{\vartheta}}\sin\vartheta}\frac{d(\sqrt{\Delta_{\vartheta}}\sin\vartheta)}{d\vartheta},\qquad
H(\vartheta)=a\omega\sin\vartheta-\frac{\widehat{k}}{\sin\vartheta}, \label{4}\\
\fl
&&\mathcal{L}^{\dag}_{1/2}=\displaystyle\frac{d}{d\vartheta}-\frac{\Xi
H(\vartheta)}{\Delta_{\vartheta}}+
\frac{1}{2\sqrt{\Delta_{\vartheta}}\sin\vartheta}\frac{d(\sqrt{\Delta_{\vartheta}}\sin\vartheta)}{d\vartheta}.\label{5}
\end{eqnarray}
By means of the further ansatz
\begin{eqnarray*}
\fl F_1(r,\vartheta)&=&R_{-}(r)S_{-}(\vartheta),\qquad
F_2(r,\vartheta)=R_{+}(r)S_{+}(\vartheta),\\
\fl G_1(r,\vartheta)&=&R_{+}(r)S_{-}(\vartheta),\qquad
    G_2(r,\vartheta)=R_{-}(r)S_{+}(\vartheta)
\end{eqnarray*}
the Dirac equation decouples into the following systems of linear
first order differential equations for the radial $R_{\pm}$ and
angular components $S_{\pm}$ of the spinor $\Psi$
\begin{eqnarray}
\fl
&&\left( \begin{array}{cc}
     \mathcal{D}_{0}&-\rmi m_{e}r-\lambda\\
     \rmi
     m_{e}r-\lambda&\Delta_{r}\mathcal{D}^{\dag}_{1/2}\\
           \end{array} \right)\left( \begin{array}{cc}
                                     R_{-} \\
                                     R_{+}
                                     \end{array}\right)=0, \label{radial}\\
\fl
&&\left( \begin{array}{cc}
     \lambda-am_{e}\cos\theta & \sqrt{\Delta_\vartheta}~\mathcal{L}_{1/2}\\
     \sqrt{\Delta_\vartheta}~\mathcal{L}^{\dag}_{1/2}& -(\lambda+am_{e}\cos\theta)
           \end{array} \right)\left( \begin{array}{cc}
                                     S_{-} \\
                                     S_{+}
                                     \end{array}\right)=0 \label{angular}
\end{eqnarray}
where $\lambda$ is a separation constant. In the case $\Lambda=a=0$
the components $S_{\pm}$ of the angular eigenfunctions are
spin-weighted spherical harmonics \cite{pen1,gold} whereas for
$\Lambda=0$ and $a\neq 0$ the radial and angular eigenfunctions
satisfy a generalized Heun equation \cite{batic,uns}. In the next
section we show that in the more general case $\Lambda\neq 0\neq a$
the functions $R_{\pm}$ and $S_{\pm}$ satisfy the hypergeneralized
Heun equation (\ref{HHE}).

\section{Reduction to the HHE}\label{sec:3}

By means of a suitable transformation we show that the radial and
angular eigenfunctions satisfy a hypergeneralized Heun equation.
Decoupling the radial system (\ref{radial}) in favor of $R_{-}$ we
get
\begin{equation}\label{Rad}
\fl
\left[\Delta_{r}\mathcal{D}^{\dag}_{1/2}\mathcal{D}_{0}-\frac{\rmi
m_{e}\Delta_{r}}{\lambda+\rmi m_{e}r}\mathcal{D}_{0}-
(\lambda^{2}+m^{2}_{e}r^{2})\right]R_{-}=0.
\end{equation}
Analogously, if we decide to eliminate $S_{+}$ in (\ref{angular}) we
obtain
\begin{equation}\label{AN}
\fl
\left[\sqrt{\Delta_{\vartheta}}~\mathcal{L}_{1/2}\sqrt{\Delta_\vartheta}~\mathcal{L}^{\dag}_{1/2}+
\frac{am_{e}\sin\vartheta~\Delta_{\vartheta}}{\lambda+am_{e}\cos\vartheta}~\mathcal{L}^{\dag}_{1/2}+\lambda^{2}+
a^{2}m^{2}_{e}\cos^{2}\vartheta\right]S_{-}=0.
\end{equation}
\subsection{The radial equation (\ref{Rad})}
Let $r_{1},\cdots,r_{4}\in\mathbb{C}$ be the roots of the polynomial
equation $\Delta_{r}=0$. We suppose that $r_i\neq r_j$ for each
$i\neq j$ with $i,j=1,\cdots,4$. Moreover, let us introduce the
notation $r_{ij}:=r_{i}-r_{j}$. For the reduction to the HHE it is
convenient to make the variable transformation
\[
\fl z=\displaystyle\frac{r-r_{1}}{r_{2}-r_{1}}
\]
mapping $r_1$ to zero and $r_2$ to one. This allows to write the
polynomial $\Delta_r$ in a very compact form as follows
\[
\fl \Delta_{r}=r^{4}_{21}\Delta_{z},\qquad
\Delta_{z}=-\frac{\Lambda}{3}z(z-1)(z-z_{3})(z-z_{4}),\qquad
z_3=\frac{r_{31}}{r_{21}},\qquad z_4=\frac{r_{41}}{r_{21}}.
\]
Finally, taking into account that $\mathcal{D}_{0}$,
$\mathcal{D}^{\dag}_{1/2}$ are given by (\ref{1}) and (\ref{3})
equation (\ref{Rad}) becomes
\begin{equation}\label{quasi}
\fl \frac{d^{2}R_{- }}{dz^{2}}+P(z)\frac{dR_{-}}{dz}+Q(z)R_{-}=0
\end{equation}
where for future convenience we write $P(z)$ and $Q(z)$ in terms of
partial fractions as follows
\begin{eqnarray*}
\fl P(z)&=&\frac{1}{2}\left(\frac{1}{z}+
\frac{1}{z-1}+\frac{1}{z-z_{3}}+\frac{1}{z-z_{4}}\right)-\frac{1}{z-z_{5}},\qquad
z_5=\frac{r_{51}}{r_{21}},\qquad r_5:=\rmi\frac{\lambda}{m_e},\\
\fl Q(z)&=&
\frac{B_{1}}{z^{2}}+\frac{B_{2}}{(z-1)^{2}}+\frac{B_{3}}{(z-z_{3})^{2}}+\frac{B_{4}}{(z-z_{4})^{2}}+
\frac{T_{1}}{z}+\frac{T_{2}}{z-1}+\frac{T_{3}}{z-z_{3}}+\frac{T_{4}}{z-z_{4}}+\frac{T_{5}}{z-z_{5}}
\end{eqnarray*}
with
\begin{eqnarray*}
\fl&&B_{1}=-\frac{3\rmi\Xi K(r_{1})}{2\Lambda
r_{21}r_{31}r_{41}},\qquad
\hspace{2.3cm}B_{2}=-\frac{3\rmi\Xi\left[K(r_{1})-\omega(r^{2}_{2}-r^{2}_{1})\right]}{2\Lambda r_{12}r_{32}r_{42}},\\
\fl&&B_{3}=-\frac{3\rmi\Xi\left[K(r_{1})-\omega(r^{2}_{3}-r^{2}_{1})\right]}{2\Lambda
r_{13}r_{23}r_{43}},\qquad
B_{4}=-\frac{3\rmi\Xi\left[K(r_{1})-\omega(r^{2}_{4}-r^{2}_{1})\right]}{2\Lambda r_{14}r_{24}r_{34}},\\
\fl&&T_{1}=\frac{3}{\Lambda r^{4}_{21}r_{31}r_{41}r_{51}}\left\{\Xi^{2}r_{51}K(r_{1})+\rmi\Xi r^{4}_{21}
\left[K(r_{1})-\omega r_{1}r_{51}\right]-r^{4}_{21}r_{51}(\lambda^{2}+m^{2}_{e}r^{4}_{1})\right\},\\
\fl&&T_{2}=\frac{3}{\Lambda r^{4}_{21}r_{23}r_{24}r_{25}}\left\{\Xi^{2}r_{52}\left[K(r_{1})
-\omega r_{21}(r_{2}+r_{1})\right]-\rmi\Xi r^{4}_{21}\left[K(r_{1})-\omega(r_{5}r_{2}-r^{2}_{1})\right]+\right.\\
\fl&&\hspace{9.8cm}\left.-r^{4}_{21}r_{52}\left[\lambda^{2}+m^{2}_{e}(r^{2}_{1}+r_{21})^{2}\right]\right\},\\
\fl&&T_{3}=\frac{3}{\Lambda r^{3}_{21}r_{31}r_{32}r_{34}r_{35}}\left\{\Xi^{2}r_{53}\left[K(r_{1})
-\omega r_{31}(r_{3}+r_{1})\right]+\rmi\Xi r^{4}_{21}\left[K(r_{1})-\omega(r_{5}r_{3}-r^{2}_{1})\right]+\right.\\
\fl&&\hspace{9.8cm}\left.-r^{4}_{21}r_{53}\left[\lambda^{2}+m^{2}_{e}(r^{2}_{1}+r_{31})^{2}\right]\right\},\\
\fl&&T_{4}=\frac{3}{\Lambda r^{3}_{21}r_{41}r_{42}r_{43}r_{54}}\left\{\Xi^{2}r_{45}\left[K(r_{1})
-\omega r_{41}(r_{4}+r_{1})\right]-\rmi\Xi r^{4}_{21}\left[K(r_{1})-\omega(r_{5}r_{4}-r^{2}_{1})\right]+\right.\\
\fl&&\hspace{9.8cm}\left.-r^{4}_{21}r_{45}\left[\lambda^{2}+m^{2}_{e}(r^{2}_{1}+r_{41})^{2}\right]\right\},\\
\fl&&T_{5}=\frac{3\rmi\Xi r_{21}\left[K(r_{1})-\omega
r_{51}(r_{5}+r_{1})\right]}{\Lambda r_{51}r_{52}r_{53}r_{54}}.
\end{eqnarray*}
It can be checked that $T_{1}+T_{2}+T_{3}+T_{4}+T_{5}=0$. This
property of the coefficients $T_i$ will ensure at the end that the
point at infinity is a regular singular point. Finally, by
transforming $R_{-}$ according to
\[
\fl
R_{-}(z)=z^{\alpha_{1}}(z-1)^{\alpha_{2}}(z-z_{3})^{\alpha_{3}}(z-z_{4})^{\alpha_{4}}\widehat{R}_{-}(z)
\]
and requiring that $2\alpha^{2}_{i}-\alpha_{i}+2B_{i}=0$ for each
$i=1,\cdots,4$ equation (\ref{quasi}) becomes
\begin{equation}\label{ultima}
\fl
\frac{d^{2}\widehat{R}_{-}}{dz^{2}}+\widetilde{P}(z)\frac{d\widehat{R}_{-}}{dz}+\widetilde{Q}(z)\widehat{R}_{-}=0
\end{equation}
with
\begin{eqnarray*}
\fl&&\widetilde{P}(z)=\frac{1+4\alpha_{1}}{2z}+\frac{1+4\alpha_{2}}{2(z-1)}+\frac{1+4\alpha_{3}}{2(z-z_{3})}+\frac{1+4\alpha_{4}}{2(z-z_{4})}-\frac{1}{z-z_{5}},\\
\fl&&\widetilde{Q}(z)=\frac{C_{1}}{z}+\frac{C_{2}}{z-1}+\frac{C_{3}}{z-z_{3}}+\frac{C_{4}}{z-z_{4}}+\frac{C_{5}}{z-z_{5}}.
\end{eqnarray*}
where
\begin{eqnarray*}
\fl
C_1&=&\frac{\alpha_1}{z_5}-\frac{\alpha_1+4\alpha_1\alpha_4+\alpha_4}{2z_4}+
\frac{(2T_1-\alpha_1-4\alpha_1\alpha_2-\alpha_2)z_3-(\alpha_1+4\alpha_1\alpha_3+\alpha_3)}{2z_3},\\
\fl
C_2&=&\frac{\alpha_2}{z_5-1}-\frac{\alpha_2+4\alpha_2\alpha_4+\alpha_4}{2(z_4-1)}+
\frac{(2T_2+\alpha_1+4\alpha_1\alpha_2+\alpha_2)(z_3-1)-(\alpha_2+4\alpha_2\alpha_3+\alpha_3)}{2(z_3-1)},\\
\fl
C_3&=&T_3-\frac{\alpha_3}{z_3-z_5}+\frac{\alpha_1+4\alpha_1\alpha_3+\alpha_3}{2z_3}+
\frac{\alpha_2+4\alpha_2\alpha_3+\alpha_3}{2(z_3-1)}+\frac{\alpha_3+4\alpha_3\alpha_4+\alpha_4}{2(z_3-z_4)},\\
\fl
C_4&=&T_4-\frac{\alpha_4}{z_4-z_5}+\frac{\alpha_1+4\alpha_1\alpha_4+\alpha_4}{2z_4}+
\frac{\alpha_2+4\alpha_2\alpha_4+\alpha_4}{2(z_4-1)}+\frac{\alpha_3+4\alpha_3\alpha_4+\alpha_4}{2(z_4-z_3)},\\
\fl C_5&=&T_5-\frac{\alpha_1}{z_5}-\frac{\alpha_2}{z_5-1}-
\frac{\alpha_3}{z_5-z_3}-\frac{\alpha_4}{z_5-z_4}.
\end{eqnarray*}
Since
\[
\fl \sum_{n=1}^{5}C_{n}=T_{1}+T_{2}+T_{3}+T_{4}+T_{5}=0
\]
equation (\ref{ultima}) reduces to the hypergeneralized Heun
equation (\ref{HHE}).
\subsection{The angular equation (\ref{AN})}
Let $\alpha:=\Lambda a^{2}/3$ and $\beta:=\lambda/(am_e)$. If we
introduce the variable transformation $z=(1+\cos\vartheta)/2$
equation (\ref{AN}) becomes
\[
\fl \frac{d^{2}S_{-}}{dz^{2}}
+\left[\frac{\Delta^{'}_{z}}{\Delta_{z}}+\frac{2z-1}{z(z-1)}-
\frac{1}{z-z_{5}}\right]\frac{dS_{-}}{dz}
-\frac{1}{z(z-1)\Delta_{z}}\left\{-\Xi\left[a\omega(2z-1)+\frac{\widehat{k}(2z-1)}{4z(z-1)}\right]+\right.
\]
\[
\fl +\frac{\Xi\Delta^{'}_{z}}{4\Delta_{z}}\left[4a\omega
z(z-1)+\widehat{k}\right]
-\frac{1}{4}z(z-1)\Delta^{''}_{z}-\frac{3}{8}\Delta^{'}_{z}+
\frac{1}{8}z(z-1)\frac{\left.\Delta^{'}_{z}\right.^{2}}{\Delta_{z}}+
\frac{(2z-1)^{2}\Delta_{z}}{16z(z-1)}+
\]
\[
\fl +\frac{\Xi^{2}}{\Delta_{z}}\left[4a^{2}\omega^{2}z(z-1)+
\frac{\widehat{k}^{2}}{4z(z-1)}+2a\omega\widehat{k}\right]+\frac{\left.\Delta^{'}_{z}\right.^{2}}{64\Delta_{z}}+
\frac{\Xi}{2(z-z_{5})}\left[4a\omega z(z-1)+\widehat{k}\right]+
\]
\[
\fl
\left.+\frac{z(z-1)\Delta^{'}_{z}}{4(z-z_{5})}+\frac{(2z-1)\Delta_{z}}{4(z-z_{5})}-\frac{1}{2}\Delta_{z}+\lambda^{2}+
a^{2}m^{2}_{e}(2z-1)^{2}\right\}S_{-}=0
\]
with
\[
\fl \Delta_{z}=4\alpha(z-z_3)(z-z_4),\qquad
z_{3}=\frac{1}{2}-\frac{1}{2\sqrt{-\alpha}},\qquad
z_{4}=\frac{1}{2}+\frac{1}{2\sqrt{-\alpha}},\qquad
z_5=\frac{1-\beta}{2}.
\]
The above differential equation can be written in a more amenable
form if we make a partial fraction expansion of its coefficient
functions. Thus, we obtain
\begin{equation}\label{rumba}
 \fl
\frac{d^{2}S_{-}}{dz^{2}}+\mathcal{P}(z)\frac{dS_{-}}{dz}+\mathcal{Q}(z)S_{-}=0
\end{equation}
with
\begin{eqnarray*}
\fl
P(z)&=&\frac{1}{z}+\frac{1}{z-1}+\frac{1}{z-z_{3}}+\frac{1}{z-z_{4}}-\frac{1}{z-z_{5}},\\
\fl
Q(z)&=&\frac{\mathcal{B}_{1}}{z^{2}}+\frac{\mathcal{B}_{2}}{(z-1)^{2}}+\frac{\mathcal{B}_{3}}{(z-z_{3})^{2}}
+\frac{\mathcal{B}_{4}}{(z-z_{4})^{2}}
+\frac{\mathcal{T}_{1}}{z}+\frac{\mathcal{T}_{2}}{z-1}+\frac{\mathcal{T}_{3}}{z-z_{3}}+\frac{\mathcal{T}_{4}}{z-z_{4}}
+\frac{\mathcal{T}_{5}}{z-z_{5}}
\end{eqnarray*}
where
\begin{eqnarray*}
\fl
B_{1}&=&-\frac{(\alpha+1-2\widehat{k}~\Xi)^2}{16(\alpha+1)^2},\qquad
B_{2}=-\frac{(\alpha+1+2\widehat{k}~\Xi)^2}{16(\alpha+1)^2},\\
\fl
B_{3}&=&\frac{1}{64\alpha\sqrt{-\alpha}(\alpha+1)^{2}}\left[16\Xi^{2}\sqrt{-\alpha}\Psi^{2}_{\alpha}-16\Xi\alpha(1+\alpha)\Psi_{\alpha}-\alpha(1+\alpha)^{2}\right],\quad
\Psi_{\alpha}:=\left[a\omega(1+\alpha)-ka\right],\\
\fl
B_{4}&=&B_{3}+\frac{1}{32\sqrt{-\alpha}(1+\alpha)}\left[(1+\alpha)(1+16\Xi
a\omega)-16\Xi ka\right],
\end{eqnarray*}
\[
\fl
\mathcal{T}_{1}=\tau_{0}+\tau_1+\frac{24z_5(z_4-1)-4z_4^2(6z_5-4)}{64z_5z_4^2}
+\frac{8z_5(\lambda^2+a^2m^2_e)+\alpha z_5(1+12z_4)+4\Xi(2z_5
a\omega-\widehat{k})}{32\alpha z_3 z_4 z_5}
\]
\[
\fl
\mathcal{T}_2=\tau_2+\tau_3+\frac{8z_4^2(3z_5-1)+8z_4(5-9z_5)+47z_5-31}{64(z_5-1)(z_4-1)^2)}+
\]
\[
+\frac{(1-z_5)[\alpha(12z_4-11)+8(\lambda^2+a^2m_e^2-a\omega\Xi)]+4\Xi\widehat{k}}
{32\alpha(z_3-1)(z_4-1)(z_5-1)},
\]
\[
\fl \mathcal{T}_3=\tau_0-\tau_1+\tau_2-\tau_3+\tau_4+\tau_5^{-}
-\frac{\Xi^2\widehat{k}(2z_4-1)[4a\omega
z_4(z_4-1)+\widehat{k}]}{32\alpha^2 z_4^3(z_4-z_3)^2(z_4-1)^3}+
\]
\[
 -\frac{z_5[8(\lambda^2+a^2m^2_e+\Xi
a\omega)+\alpha(1+12z_4)]-4\Xi\widehat{k}}{32\alpha z_3z_4z_5} +
\]
\[
-\frac{(z_5-1)[8(\lambda^2+a^2m^2_e-\Xi
a\omega)+\alpha(12z_4-11)]-4\Xi\widehat{k}}{32\alpha
(z_3-1)(z_4-1)(z_5-1)}+
\]
\[
\fl
+\frac{(z_4-z_5)[8(\lambda^2+a^2m_e^2)+\alpha+8z_4(1-z4)(\alpha-4a^2m_e^2)]-8\Xi
a\omega(z_4+z_5+2z_4z_5)+4\Xi\widehat{k}}{32\alpha
z_4(z_4-1)(z_4-z_3)(z_4-z_5)},
\]
\[
\fl
\mathcal{T}_{4}=\tau_{5}^{+}-\frac{16z_4^4-56z_4^3+24z_4^2z_5+38z_4^2-22z_4z_5+z_4-z_5}{64z_4^2(z_4-z_5)(z_4-1)^2}
+
\]
\[
\fl
+\frac{8(z_4-z_5)[\lambda^2+a^2m_e^2(2z_4-1)^2]+\alpha(z_4-1)(1+8z_4z_5-8z_4^2)-8\Xi
a\omega(z_4+z_5-2z_4z_5)+4\Xi\widehat{k}}{32\alpha
z_4(z_4-1)(z_4-z_5)(z_3-z_4)}
\]
\[+\frac{\Xi^2\widehat{k}[4a\omega
z_4(z_4-1)(2z_4-1)+\widehat{k}(2z_4-1)]}{32\alpha^2z_4^3(z_4-z_3)^2(z_4-1)^3}.
\]
Moreover,
\[
\fl \tau_{0}:=-\frac{\Xi^2\widehat{k}^2}{32\alpha^2z_4^2z_3^3},\quad
\tau_{1}:=\frac{z_4(\alpha^2
z_4^2+8\Xi^2a\omega\widehat{k})-2\Xi^2\widehat{k}^2(1+z_4)}{64\alpha^2z_3^2z_4^3},\quad
\tau_2:=\frac{\Xi^2\widehat{k}^2}{32\alpha^2(z_3-1)^3(z_4-1)^2},
\]
\[
\fl
\tau_3:=\frac{(1-z_4)[\alpha^2(1-z_4)^2+2\Xi^2\widehat{k}(4a\omega-\widehat{k})]-2\Xi^2\widehat{k}^2}
{64\alpha^2(z_3-1)^2(z_4-1)^3},
\]
\[
\fl \tau_4:=\frac{2z_5(1-z_5)[\alpha(z_5-z_4)+2\Xi
a\omega]-\Xi\widehat{k}}{8\alpha z_5(1-z_5)(z_3-z_5)(z_4-z_5)},\quad
\tau_{5}^{\pm}:=\frac{\Xi^2\{8a\omega
z_4(1-z_4)[2a\omega(1-z_4)\pm\widehat{k}]+\widehat{k}^2\}}{32\alpha^2z_4^2(z_3-z_4)^3(1-z_4)^2}.
\]
Also in this case we have that
$\mathcal{T}_{1}+\mathcal{T}_{2}+\mathcal{T}_{3}+\mathcal{T}_{4}+\mathcal{T}_{5}=0$
which ensures that the point at infinity is a regular singular
point. Finally, if we transform $S_{-}$ according to
\[
\fl
S_{-}(z)=z^{\gamma_{1}}(z-1)^{\gamma_{2}}(z-z_{3})^{\gamma_{3}}(z-z_{4})^{\gamma_{4}}\widehat{S}_{-}(z)
\]
together with the requirement $\gamma^{2}_{i}+B_{i}=0$ for each
$i=1,\cdots,4$ equation (\ref{rumba}) becomes
\begin{equation}\label{rumbaf}
\fl
\frac{d^{2}\widehat{S}_{-}}{dz^{2}}+\widetilde{\mathcal{P}}(z)\frac{d\widehat{S}_{-}}{dz}+\widetilde{\mathcal{Q}}(z)\widehat{S}_{-}=0
\end{equation}
with
\begin{eqnarray*}
\fl
&&\widetilde{P}(z)=\frac{1+2\gamma_{1}}{z}+\frac{1+2\gamma_{2}}{z-1}+\frac{1+2\gamma_{3}}{z-z_{3}}
+\frac{1+2\gamma_{4}}{z-z_{4}}-\frac{1}{z-z_{5}},\\
\fl
&&\widetilde{Q}(z)=\frac{\widetilde{C}_{1}}{z}+\frac{\widetilde{C}_{2}}{z-1}
+\frac{\widetilde{C}_{3}}{z-z_{3}}+\frac{\widetilde{C}_{4}}{z-z_{4}}+\frac{\widetilde{C}_{5}}{z-z_{5}}
\end{eqnarray*}
where
\begin{eqnarray*}
\fl
\widetilde{C}_1&=&\frac{\gamma_1}{z_5}-\frac{\gamma_1+2\gamma_1\gamma_4+\gamma_4}{z_4}+
\frac{(\mathcal{T}_1-\gamma_1-2\gamma_1\gamma_2-\gamma_2)z_3-(\gamma_1+2\gamma_1\gamma_3+\gamma_3)}{z_3},\\
\fl
\widetilde{C}_2&=&\frac{\gamma_2}{z_5-1}-\frac{\gamma_2+2\gamma_2\gamma_4+\gamma_4}{z_4-1}+
\frac{(\mathcal{T}_2+\gamma_1+2\gamma_1\gamma_2+\gamma_2)(z_3-1)-(\gamma_2+2\gamma_2\gamma_3+\gamma_3)}{z_3-1},\\
\fl
\widetilde{C}_3&=&\mathcal{T}_3-\frac{\gamma_3}{z_3-z_5}+\frac{\gamma_1+2\gamma_1\gamma_3+\gamma_3}{z_3}+
\frac{\gamma_2+2\gamma_2\gamma_3+\gamma_3}{z_3-1}+\frac{\gamma_3+2\gamma_3\gamma_4+\gamma_4}{z_3-z_4},\\
\fl
\widetilde{C}_4&=&\mathcal{T}_4-\frac{\gamma_4}{z_4-z_5}+\frac{\gamma_1+2\gamma_1\gamma_4+\gamma_4}{z_4}+
\frac{\gamma_2+2\gamma_2\gamma_4+\gamma_4}{z_4-1}+\frac{\gamma_3+2\gamma_3\gamma_4+\gamma_4}{z_4-z_3},\\
\fl
\widetilde{C}_5&=&\mathcal{T}_5-\frac{\gamma_1}{z_5}-\frac{\gamma_2}{z_5-1}-
\frac{\gamma_3}{z_5-z_3}-\frac{\gamma_4}{z_5-z_4}.
\end{eqnarray*}
Finally, since
\[
\fl
\sum_{n=1}^{5}\mathcal{C}_{n}=\mathcal{T}_{1}+\mathcal{T}_{2}+\mathcal{T}_{3}+\mathcal{T}_{4}+\mathcal{T}_{5}=0
\]
equation (\ref{rumbaf}) reduces to the hypergeneralized Heun
equation (\ref{HHE}).

\section{Conclusions}
In this paper we have related the radial and angular Dirac equations
for a massive fermion in the Kerr-Newman-deSitter geometry to a HHE.
In view of the results obtained in \cite{batic} since the
Kerr-Newman-deSitter metric becomes the Kerr-Newman metric in the
limit of a vanishing cosmological constant it follows that the
generalized Heun equation is a confluent case of the HHE for
$\Lambda\to 0$. Finally, since it has been proved in \cite{Kamran}
that the neutrino equation can be separated in the class
$\mathcal{D}_0$ of algebraically special Petrov type-D vacuum
solutions with cosmological constant whereas the massive Dirac
equation can be separated only in the Carter's subclass
$\widetilde{A}$ our result rises the question if the separated
neutrino and massive Dirac equations can be also reduced to a HHE in
such more general metrics. We reserve the study of this problem for
future investigations.

\end{document}